\documentclass[RNAAS]{aastex63}
\usepackage{amsmath}

\graphicspath{{./}{figures/}}

\definecolor{gcolor}{RGB}{51, 153, 51}
\definecolor{rcolor}{RGB}{211, 33, 45}

\newcommand{\code}[1]{\texttt{\detokenize{#1}}}

\usepackage{color}

\newcommand{\vplanet}{\texttt{VPLanet}}

\newcommand{\appos}{\texttt{approxposterior}}

\newcommand{\D}{\mathcal{D}}

\newcommand{\msun}{M$_{\odot}$}
\newcommand{\lsun}{L$_{\odot}$}

\newcommand{\mstar}{$m_{*}$}
\newcommand{\tsat}{$t_{sat}$}
\newcommand{\fsat}{f$_{sat}$}
\newcommand{\age}{\rm age}
\newcommand{\bxuv}{$\beta_{\rm XUV}$}
\newcommand{\lxuv}{L$_{\rm XUV}$}
\newcommand{\lbol}{L$_{\rm bol}$}
\newcommand{\lratio}{L$_{\rm XUV}$/L$_{\rm bol}$}

\begin{document}
\title{Improved Constraints for the XUV Luminosity Evolution of Trappist-1}

\correspondingauthor{Jessica Birky}
\email{jbirky@uw.edu}

\author[0000-0002-7961-6881]{Jessica Birky}
\affil{Astronomy Department, University of Washington, Seattle, WA 98195, USA}

\author[0000-0001-6487-5445]{Rory Barnes}
\affil{Astronomy Department, University of Washington, Seattle, WA 98195, USA}
\affil{NASA Virtual Planetary Laboratory Lead Team, USA}

\author[0000-0001-9293-4043]{David P. Fleming}
\affil{Astronomy Department, University of Washington, Seattle, WA 98195, USA}

\begin{abstract}
    We re-examine the XUV luminosity evolution of TRAPPIST-1 utilizing new observational constraints (XUV and bolometric luminosity) from multi-epoch X-ray/UV photometry. Following the formalism presented in \cite{fleming_xuv_2020}, we infer that TRAPPIST-1 maintained a saturated XUV luminosity, relative to the bolometric luminosity, of $\log_{10}$(\lxuv/\lbol) $= -3.03_{-0.23}^{+0.25}$ at early times for a period of \tsat $= 3.14_{-1.46}^{+2.22}$ Gyr. After the saturation phase, we find \lxuv\ decayed over time by an exponential rate of \bxuv $= -1.17_{-0.28}^{+0.27}$. Compared to our inferred age of the system, $\age = 7.96_{-1.87}^{+1.78}$ Gyr, our result for \tsat\ suggests that there is only a $\sim4\%$~chance that TRAPPIST-1 still remains in the saturated phase today, which is significantly lower than the previous estimate of 40\%. Despite this reduction in \tsat, our results remain consistent in the conclusion that the TRAPPIST-1 planets likely received an extreme amount XUV energy---an estimated integrated XUV energy of $\sim10^{30}-10^{32}$ erg over the star's lifetime. 
\end{abstract}

% =========================================================
\section{Introduction} 

X-ray/extreme ultraviolet (XUV; $\sim1-1000\,{\rm A}$) luminosity is a fundamental aspect of stars, and a key driver for influencing the atmospheric retention and composition of potentially habitable exoplanets \citep{Segura10}. One star of particular interest in this regard is TRAPPIST-1, a very late M dwarf orbited by at least 7 planets \citep{Gillon16}, with 3--5 potentially habitable today.

A recent study by \citet[][hereafter F20]{fleming_xuv_2020} inferred the range of evolutionary histories of this star permitted by observations \citep{wheatley_strong_2017} and the \cite{Ribas05} empirical model of XUV evolution (Equation \ref{eqn:ribas}). F20 found that the star is likely still active, with XUV energy representing about 0.1\% of the total bolometric luminosity.
After publication, two new data sets \citep{ducrot_trappist-1_2020,becker_coupled_2020} have become available, allowing for revised modeling of the star. 

Very late M dwarfs like TRAPPIST-1 are expected to possess an extended initial period of high XUV emission called the ``saturated phase''. However, the details of the evolution from saturated phase to more quiescent emission at later times is poorly understood for M dwarfs. Therefore we employ an empirical model derived from observations of FGK stars in which XUV emission (driven by magnetic activity) remains constant relative to the bolometric luminosity for a ``saturation time,'' \tsat, and then decreases exponentially afterwards \citep{Ribas05}:
\begin{align}
\frac{L_{\rm XUV}}{L_{\rm bol}} = \left\{
				\begin{array}{lcr}
					f_\mathrm{sat} &\ & t \leq t_\mathrm{sat} \\
					f_\mathrm{sat}\left(\frac{t}{t_\mathrm{sat}}\right)^{-\beta_\mathrm{XUV}} &\ & t > t_\mathrm{sat}
				\end{array}
				\right.,
\label{eqn:ribas}
\end{align}
where \lbol\ is the bolometric luminosity [\lsun], \lxuv\ is the XUV luminosity [\lsun], $f_\mathrm{sat} = \log_{10}$(\lratio) is the saturation ratio, \tsat\ is the duration of saturation phase [Gyr], \bxuv\ is the exponential decay rate of \lratio\ after saturation, and $t$ is the evolution time (age) [Gyr]. Although limited empirical analysis has been done to explicitly constrain \tsat\ for the lowest mass stars, this model is broadly consistent with measurements of Rossby numbers ($R_o = P_{rot}/\tau$, where $\tau$ is the convective turnover timescale, and rotation period is used as a proxy for age; \citealt{Pizzolato03}).  \\

% =========================================================
\section{Methods} \label{sec:methods}

Following F20, we use the open-source stellar and planetary system evolution code \vplanet\ \citep{barnes_vplanet_2020} to model stellar properties $\D$ as a function of input parameters \textbf{x}. Specifically we use the \code{STELLAR} module to interpolate evolutionary model grids from \cite{Baraffe15} to compute bolometric luminosity and Equation \ref{eqn:ribas} to compute the XUV luminosity. The posterior probability is computed as $\ln P(x | \D) \propto \ln P(\D | x) + \ln P(x)$, where the input free parameters are \textbf{x} = \{\mstar , \fsat, \tsat, age, \bxuv\} and \mstar\ denotes the mass [\msun]. We keep the prior $\ln P(x) $ consistent with the F20 assumptions, as illustrated in red in Figure \ref{fig:1}. We use a Gaussian likelihood $\ln P(\D | x)$ with parameters $\D$ = \{\lbol, \lxuv \}, where the bolometric and XUV luminosity of the host star evaluated at the present-day age of the system. The adopted values for the likelihood function are $L_{\rm XUV} = (1.77 \pm 0.22)\times10^{-7} \,$\lsun\ from \cite{becker_coupled_2020} and $L_{\rm bol} = (5.53 \pm 0.19)\times10^{-4} \,$\lsun\ from \cite{ducrot_trappist-1_2020}. 

For computational expedience, we sample the posterior using the approximate Markov chain Monte Carlo (MCMC) framework \appos\ \citep{fleming_approxposterior_2018}, which uses a Gaussian process (GP) surrogate model to estimate $P(x | \D)$ from training samples of \textbf{x}. 
Similar to the procedure of F20, we trained the GP on an initial set of 50 \vplanet\ simulations (randomly distributed over the prior space of \textbf{x}). We then iteratively added training sample points, checked the GP convergence after every addition of 100 active-learning sampled training points, and found that the algorithm converged after 23 iterations, or a total of 2,350 training points (see the Appendix of F20 for further details on \appos\ configuration, sampling procedure, and convergence criteria). The code and full posterior samples are available on github\footnote{\href{http://doi.org/10.5281/zenodo.4774198}{http://doi.org/10.5281/zenodo.4774198}} \citep{jessica_birky_2021_4774198}. \\

% =========================================================
\section{Results} 

\begin{figure}[h!]
    \begin{center}
    \includegraphics[width=\linewidth]{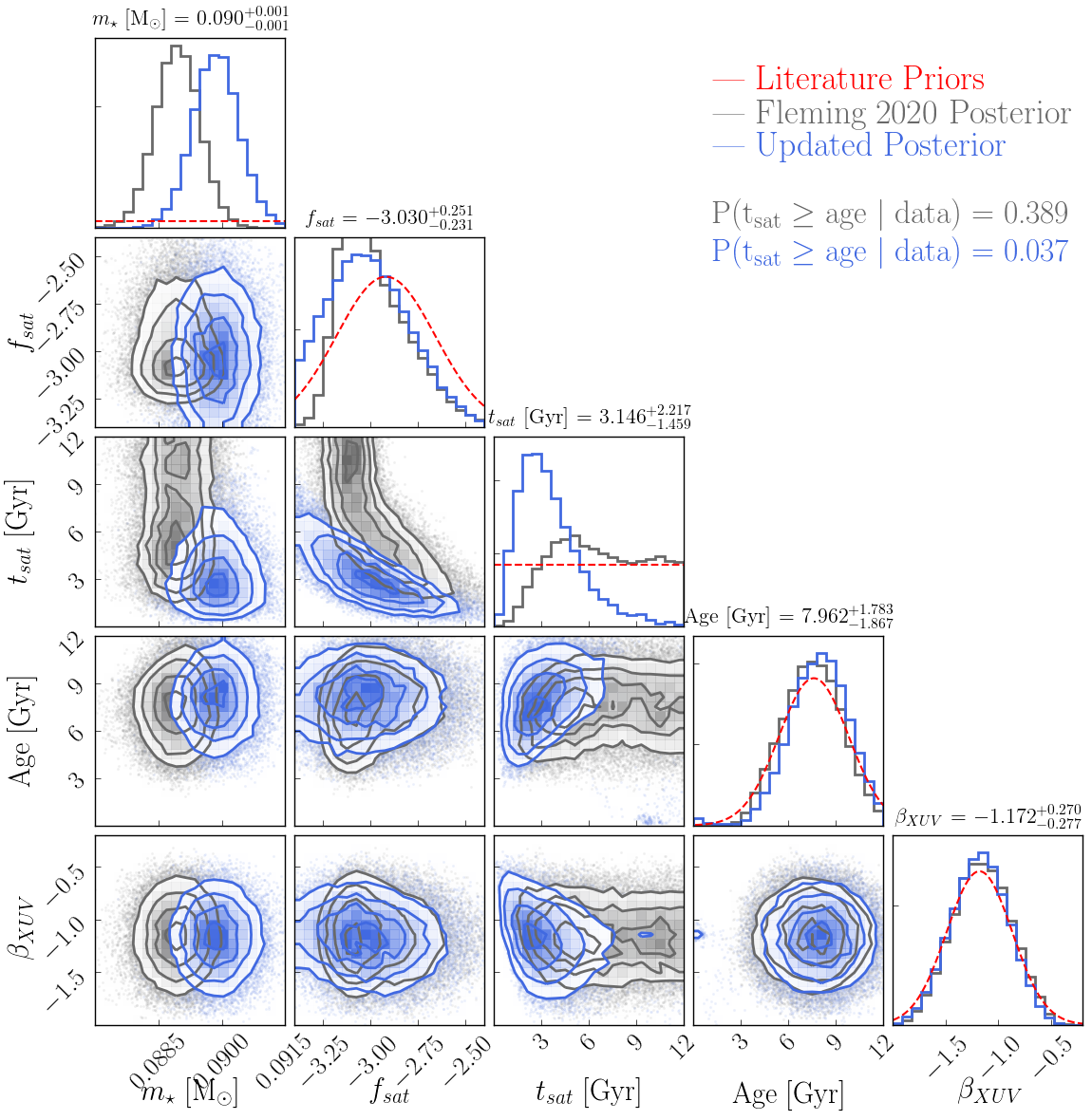}
    \caption{Posterior using the updated likelihood values described in Section \ref{sec:methods} (blue) compared to the posterior from F20 (grey). The prior distributions adopted by both works are shown in red in each histogram panel. 
    \label{fig:1}}
    \end{center}
\end{figure}

Our resulting posterior distribution is displayed in Figure \ref{fig:1}, which shows that the \fsat, \tsat, age, and \bxuv\ distributions are all constrained to better precision than in F20. Most significant is the \tsat\ distribution, which is more heavily distributed to younger values, changing the probability that TRAPPIST-1 is still saturated today from 39\% in the original result to only 3.7\% here. This significant change to \tsat\ results from adopting the \lratio\ value from \cite{becker_coupled_2020}, which is about a factor of 2 smaller than the value adopted in F20 from \cite{wheatley_strong_2017} and is likely more representative of TRAPPIST-1's quiescent state than the \cite{wheatley_strong_2017} result.

We find that the mass distribution (\mstar$ = 0.090\pm0.001$) is within a $\sim1\sigma$ agreement with the F20 result (\mstar$ = 0.089\pm0.001$). Despite using an uninformative prior, our mass result is also within a $\sim1\sigma$ agreement compared to the independent empirical mass-luminosity estimates from \cite{mann_how_2019}  (\mstar$ = 0.0898\pm0.0023$), but with an uncertainty constraint which is twice as tight. We note however that our analysis does not account for inherent uncertainties within the stellar evolution models (i.e. it assumes that the Baraffe models perfectly predict the mass-luminosity relationship of a star over time), and does not account for metallicity variation. Thus our posterior may underestimate the true uncertainty for mass.

Integrating our evolution model over the inferred age of the system for our best fit (median posterior) parameters, we estimate that TRAPPIST-1's planets received total XUV energies of $\sim \{ 2\times 10^{32}, \, 1\times 10^{32}, \, 3\times 10^{31}, \, 2\times 10^{31}, \, 2\times 10^{31}, \, 1\times 10^{31}, \, 3\times 10^{30} \}$ erg for planets b--h respectively over their lifetime. 
This reanalysis suggests the planets have received $\sim 15\%$ less XUV energy than predicted by F20. This change is modest despite the significant change in \tsat\ because most of the XUV luminosity is emitted during the pre-main sequence, and the larger mass that we infer results in higher \lbol\ and, hence, higher \lxuv.
Our updated estimates range by a factor of $\sim10-1000\times$ larger than the total XUV energy received by Earth over the Sun's lifetime, which is $\sim5\times 10^{29}$ erg (estimated by the \fsat, \tsat, and \bxuv\ values of solar-type stars from \citealt{Ribas05}). 
Thus, the primary conclusion from F20, that TRAPPIST-1's planets likely received extreme amounts of XUV radiation over their lifetime, remains unchanged.  \\ 

% =========================================================

% =========================================================
\clearpage

\bibliographystyle{aasjournal}
\bibliography{trappist,RoryBarnes}

\end{document}